\documentclass[10pt,conference]{IEEEtran}
\IEEEoverridecommandlockouts
\usepackage{cite}
\usepackage{amsmath,amssymb,amsfonts}
\usepackage{algorithmic}
\usepackage{graphicx}
\usepackage{textcomp}
\usepackage{xcolor}
\usepackage{cleveref}
\usepackage{caption}
\usepackage{enumitem}
\usepackage{tabularray}
\usepackage[most]{tcolorbox}
\usepackage[normalem]{ulem}
\usepackage{url}
\usepackage{balance}
\def\BibTeX{{\rm B\kern-.05em{\sc i\kern-.025em b}\kern-.08em
    T\kern-.1667em\lower.7ex\hbox{E}\kern-.125emX}}

\begin{document}

\title{Token Sugar: Making Source Code Sweeter for LLMs through Token-Efficient Shorthand}

\author{
    \IEEEauthorblockN{
        Zhensu Sun\IEEEauthorrefmark{2},
        Chengran Yang\IEEEauthorrefmark{2},
        Xiaoning Du\IEEEauthorrefmark{3}\IEEEauthorrefmark{1},
        Zhou Yang\IEEEauthorrefmark{4},
        Li Li\IEEEauthorrefmark{5}, and
        David Lo\IEEEauthorrefmark{2}
    }
    \IEEEauthorblockA{
        \IEEEauthorrefmark{2}Singapore Management University \quad
        \IEEEauthorrefmark{3}Monash University\quad
        \IEEEauthorrefmark{4}University of Alberta \quad
        \IEEEauthorrefmark{5}Beihang University
    }
    \IEEEauthorblockA{
        Emails: \{zssun, cryang, davidlo\}@smu.edu.sg,
        xiaoning.du@monash.edu,
        yz25@ualberta.ca,
        lilicoding@ieee.org
    }
    \IEEEauthorblockA{\IEEEauthorrefmark{1}Corresponding author}
}







\maketitle

\begin{abstract}
Large language models (LLMs) have shown exceptional performance in code generation and understanding tasks, yet their high computational costs hinder broader adoption.
One important factor is the inherent verbosity of programming languages, such as unnecessary formatting elements and lengthy boilerplate code.
This leads to inflated token counts in both input and generated outputs, which increases inference costs and slows down the generation process.
Prior work improves this through simplifying programming language grammar, reducing token usage across both code understanding and generation tasks.
However, it is confined to syntactic transformations, leaving significant opportunities for token reduction unrealized at the semantic level.

In this work, we propose \textit{Token Sugar}, a concept that replaces frequent and verbose code patterns with reversible, token-efficient shorthand in the source code.
To realize this concept in practice, we designed a systematic solution that mines high-frequency, token-heavy patterns from a code corpus, maps each to a unique shorthand, and integrates them into LLM pretraining via code transformation.
With this solution, we obtain 799 (code pattern, shorthand) pairs, which can reduce up to 15.1\% token count in the source code and is complementary to existing syntax-focused methods.
We further trained three widely used LLMs on Token Sugar-augmented data.
Experimental results show that these models not only achieve significant token savings (up to 11.2\% reduction) during generation but also maintain near-identical Pass@1 scores compared to baselines trained on unprocessed code.

\end{abstract}

\begin{IEEEkeywords}
Large Language Model, Code Simplification, AI-friendly Code Representation
\end{IEEEkeywords}

\section{Introduction}
Recent breakthroughs in large language models (LLMs) have unlocked remarkable capabilities in code-related tasks~\cite{hou2024large}, such as code generation, summarization, and translation.
LLMs process information on the basis of tokens, where inputs are tokenized into a token flow and sequentially processed to generate output tokens one after another.
Naturally, the token count in both prompts and model outputs has a direct influence on inference speed and cost.
This is also evidenced by the fact that most commercial LLM APIs charge based on the volume of tokens in inputs and outputs~\cite{AnthropicPricing2025,GoogleGeminiPricing2025}.
As a result, there is growing interest in techniques that improve the information representation efficiency of source code, i.e., reducing its token counts without compromising semantic fidelity or degrading the performance and user experience of LLMs.

Modern programming languages are inherently verbose by design.
This verbosity manifests in two principal forms: (1) functional verbosity, where necessary constructs require more tokens than theoretically needed to express the same semantics (e.g., extensive class definitions in object-oriented programming), and (2) non-functional verbosity, where elements exist solely for human readability or tooling compatibility without affecting runtime behavior (e.g., optional semicolons as statement terminators in JavaScript or Python's line continuation characters).
While such unnecessary verbosity is helpful to human readability, the additional tokens impose significant overhead for LLMs, increasing costs and slowing generation.

It motivates researchers to remove or rewrite dispensable code tokens to reduce the inference or training cost.
Existing studies~\cite{wang2025leancode,Zhang2022DietCI,wang2024natural} center on reducing input length and optimizing code understanding tasks like code summarization and code search.
Our study focuses on an important yet harder problem: \textbf{how to reduce the token usage not only for code understanding but also for code generation?}
This problem is fundamentally more challenging because code generation requires producing compilable, executable, and correct code.
Simply removing some tokens (e.g., those that models pay less attention to~\cite{Zhang2022DietCI}) would break the syntactic and semantic integrity of the generated code, which is not desired by developers or code execution environments.

The closest study that tackles this problem is by Sun et al.~\cite{sun2024ai}; they manually design a set of reversible syntactic transformations for Python and derive a new grammar called SimPy.
The model trained with SimPy can understand and generate the compact SimPy code for faster inference, which is equivalently transformed from/to human-readable Python source code, allowing human developers to work with.
However, such transformations in SimPy are inherently limited to syntactic structures, failing to address the large amount of \textit{semantic redundancies} embedded in identifiers, APIs, or design patterns.
For example, due to its lengthy naming conventions, a common API invocation like ``pandas.DataFrame.to\_dict()'' still contains considerable tokens regardless of how the syntax is simplified.
To quantify this argument, we analyzed a Github code corpus, the Python subset of starcoderdata~\cite{li2023starcoder} and found that when tokenized using GPT-4's tokenizer, only 25.5\% of tokens correspond to Python syntax elements like keywords and symbols.
This reveals a fundamental limitation of purely syntactic approaches: they leave significant token savings unrealized beyond the scope of syntax.

\begin{figure*}[!t]
    \centering
    \includegraphics[width=0.9\textwidth]{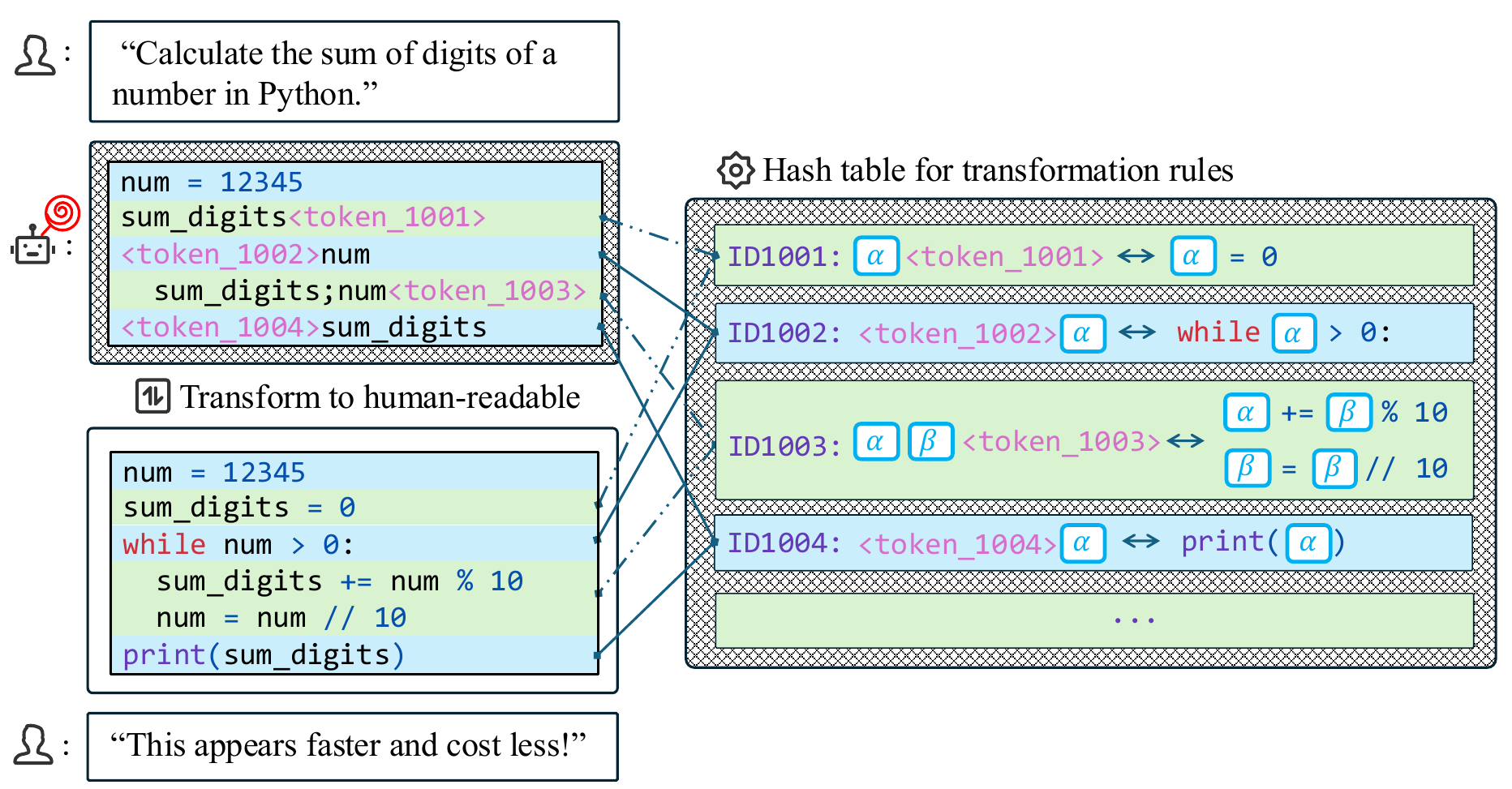}
    \caption{
    A demonstration of how Token Sugar facilitates LLMs' code generation process.
    The LLM equipped with Token Sugar can generate a compact, sugarized representation of the code using only 23 tokens, compared to the 40 tokens required by a vanilla LLM for the same logic.}
    \label{fig:demo}
\end{figure*}

To fill this gap, we propose Token Sugar.
It is designed to be bijective transformations: common code patterns are replaced with compact, token-efficient, equivalent shorthand.
Such shorthand can significantly reduce token usage during LLM inference and can be deterministically reverted to the original code.
It thus provides LLMs with a leaner token stream while preserving full semantic fidelity for downstream compilation or human interpretation.
In~\Cref{fig:demo}, we demonstrate how Token Sugar can be used.
When a code generation request is sent to a Token-Sugar-adapted LLM, the model generates the code using the shorthands of Token Sugar, consuming only 23 tokens (measured by GPT-4o's tokenizer).
Specifically, the generated code contains four different shorthands, each marked by a special token (e.g., $\langle token\_1001 \rangle$) that corresponds to a specific transformation rule.
To restore human readability, a lightweight post-processing step converts the shorthand back to standard code.
The system uses each special token's ID as an index to retrieve the corresponding transformation rule from a pre-maintained hash table.
Through this process, the compact shorthand is expanded into fully readable code that would have required 40 tokens if generated directly by a standard LLM.

However, realizing Token Sugar in practice poses several key challenges:
First, which code patterns should be sugarized?
Not all verbose patterns are equally valuable for sugarization. 
Some patterns might be too rare to justify the overhead of creating a shorthand, while others might already be relatively concise, leading to minimal benefit from sugarization.
A scalable, data-driven approach for mining frequent and token-heavy code patterns is needed.
Second, how should token sugars be represented?
The shorthand must minimize token usage while remaining unambiguous and reversible in varied code contexts, which is still unexplored in our research community.
Last but not least, can LLMs learn to use token sugars effectively?
Token sugars introduce new abstractions and vocabulary tokens.
LLMs must be trained to recognize, interpret, and generate these forms without degrading their core coding capabilities.

In this paper, we present a comprehensive framework for Token Sugar addressing these challenges.
Our approach is developed for Python as a proof-of-concept, but can be easily generalized to other programming languages with minor modifications.
At its core, Token Sugar operates by identifying frequently occurring, verbose code patterns that can be compactly represented by reversible shorthands.
These patterns are not arbitrarily chosen; instead, they are mined from real-world usage data, e.g., the code generated by an LLM application in response to its real-world users.
Specifically, we employ frequent subtree mining on generalized abstract syntax trees (ASTs) derived from such usage data.
We then filter these patterns to ensure that they are commonly used during inference, token-heavy in their current form, and sufficiently existing in training data.
This selection process enables our sugarization efforts to yield significant token savings for the LLM.
For these chosen patterns, we define a universal compact shorthand, incorporating specific delimiters and special tokens to ensure unambiguous reversibility back to the original code.
Finally, we transform the code samples that match these patterns into their corresponding shorthands, creating a sugarized dataset ideal for pretraining or continual pretraining LLMs. 
During this process, we utilize an optimization algorithm to resolve potential overlaps between different patterns, maximizing the overall token savings.
The model trained with the sugarized dataset can thus generate or understand code containing token sugars.
Notably, the sugarized code can be automatically converted back to its original, human-readable form through rule-based pre/post-processing.

To evaluate the feasibility and effectiveness of Token Sugar, we conduct a comprehensive evaluation across three different LLMs, Pythia, Qwen-2.5, and Llama-3.2.
We mine 799 token sugars from a LeetCode Python solution dataset and pre-train each LLM on a sugarized Python subset of StarCoderData~\cite{li2023starcoder}, integrating these sugars to enhance function-level code generation efficiency.
The evaluation results show that Token Sugar achieves significant token reductions of up to 15.1\%.
Furthermore, it is compatible with existing syntax-level simplification techniques.
When combined with Simpy (a state-of-the-art method that reduces tokens by 15.3\% alone), the hybrid approach yields over 22\% token savings.
Notably, LLMs trained on sugarized data retain near-identical Pass@1 performance compared to baselines (i.e., models trained on unprocessed data), indicating negligible degradation in generation quality.
Further analysis reveals that stronger models leverage token sugars more strategically, exhibiting zero desugarization failures, highlighting their robustness.
These results position Token Sugar as a practical, scalable, and effective solution for optimizing LLM efficiency.

The source code of the paper is available at \url{https://github.com/v587su/TokenSugar/}.
Our contribution can be summarized as follows:
\begin{itemize}[leftmargin=*]
\item We propose Token Sugar, a novel concept that replaces frequent and verbose code patterns with reversible, token-efficient shorthand in the source code, enabling LLMs to operate on shorter, more efficient token sequences without losing fidelity.
\item We realize Token Sugar with a practical solution, including the mining of frequent token-heavy code patterns, the design of compact and unambiguous shorthand, and an integration strategy for training LLMs on sugarized data.
\item We experimentally demonstrate that Token Sugar can significantly reduce token usage and improve code generation efficiency, with minimal impact on generation quality.
\end{itemize}

\section{Token Sugar}
While grammar-level simplifications can reduce syntactic redundancy in code~\cite{sun2024ai}, they fall short in addressing the verbosity stemming from semantics, such as lengthy API names, verbose expressions, and repeated idioms. 
To close this gap, we propose to capture frequently recurring semantic constructs, regardless of how they are expressed, and represent them with a token-efficient, reversible shorthand.
Since it makes the code ``sweeter'' for LLMs to use, we name this approach \textit{Token Sugar}.
Notably, being reversible, Token Sugar is compatible with the DualCode~\cite{sun2024ai} inference framework, which enables human users to interact with human-readable source code, while LLMs still leverage the efficiency of simplified code during the inference process.

\begin{figure}[!t]
    \centering
    \includegraphics[width=\columnwidth]{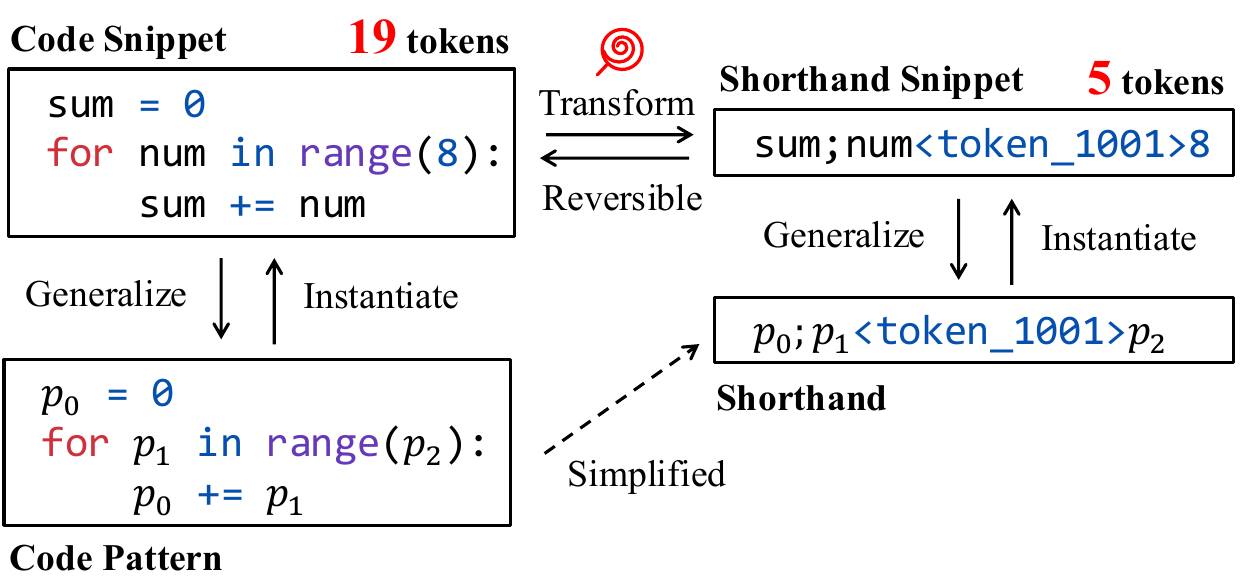}
    \caption{The conceptual relationship between the code pattern and its shorthand. The code snippet on the left costs 19 tokens measured by the tokenizer of GPT-4, but only 5 tokens when sugarized into our shorthand on the right.}
    \label{fig:example}
\end{figure}

\subsection{Problem Setup}\label{sec:problem-setup}
Token Sugar is a lossless program simplification method, where it defines a bijective transformation that common code patterns are replaced with compact, token-efficient, equivalent shorthand.
To ease the understanding, in the rest of this paper, we use Token Sugar to denote our proposed method and token sugars for the transformations.
All token sugars are built upon three components: the code pattern to be sugarized, the shorthand as the sugarized representation, and the transformation rule that defines how to convert between the code pattern and the shorthand.
We formalize a token sugar as a transformation $T: C \rightarrow S$, where $C$ is the set of code instances matching a particular code pattern, and $S$ is the set of corresponding shorthand strings.
The transformation $T$ must satisfy $T^{-1}(T(c)) = c$ for all $c \in C$, ensuring that the original code is always recoverable.
In~\Cref{fig:example}, we illustrate the conceptual relationship between code pattern, shorthand, and their instances.
In the following, we introduce them in detail.

\begin{itemize}[leftmargin=*]
    \item \textbf{Code Pattern}: 
    A code pattern is a generalized structure that captures a family of code instances through placeholders.
    Specifically, it uses placeholders to abstract over specific elements like variable names or constant values.
    When all the placeholders of a code pattern are filled with specific values, the code pattern becomes a concrete instance.
    For example, an augmented assignment pattern can be written as ``$\alpha\texttt{ += }1$'', which uses ``$\alpha$'' as the placeholder to match any elements appeared in this position.
    This code pattern can capture any instance of incrementing a variable by 1, e.g., \texttt{x += 1}, \texttt{total += 1}, etc.
    Using patterns instead of exact code allows the token sugar to be applied to a broader range of code examples with a single rule.

    \item \textbf{Shorthand}: 
    The shorthand is also a generalized structure, being the simplified counterpart of the corresponding code pattern.
    It preserves the original code pattern’s placeholders but organizes them with a streamlined structure.
    Notably, since its objective is token efficiency rather than direct execution, the structure need not comply with the language’s grammar as long as it can be unambiguously parsed back to the original form.
    For instance, the increment pattern ``$\alpha+=1$'' might be rewritten as ``$\alpha$$\langle TOKEN\rangle$'', where the special token $\langle TOKEN\rangle$ compactly encodes the ``+= 1'' operation, and the placeholder $\alpha$ remains.
    This allows LLMs to efficiently recognize and process recurring constructs while avoiding the verbosity of original patterns.

    \item \textbf{Transformation Rule}: 
    The transformation rule defines how to convert between the original code and its shorthand counterpart.
    The transformation must be bijective, i.e., each code matched by a specific code pattern has exactly one corresponding shorthand variant, and vice versa.
    During sugarization, the rule extracts placeholder values from the original code (based on the pattern) and injects them into the shorthand’s structure.
    During desugarization, it does the reverse, identifying the placeholders in the shorthand and reconstructing the original code.
    For the transformation to work reliably, the shorthand must be unambiguous: it must be clear where each placeholder starts and ends, even when embedded in other code content.
    Notably, the sugarization and desugarization process is not performed by existing parsers of the programming language.
    Instead, it relies on a converter implemented to capture the target code pattern or shorthand and perform the replacement based on these transformation rules.

\end{itemize}

\subsection{Challenges in Practice}\label{sec:challenges}
While Token Sugar offers a promising abstraction for reducing token overhead, realizing it in practice requires overcoming several key challenges.
These challenges span from selecting the right patterns, to designing usable shorthand formats, to ensuring that LLMs can effectively understand and generate sugarized code.

\begin{itemize}[leftmargin=*]
    \item \textbf{How should shorthand formats be designed?}  
    As a novel abstraction layer, the design for shorthand formats remains unexplored.
    Designing a good shorthand is not just about compressing; it must also be unambiguous and reversible while aligning well with the working mechanism of LLMs, e.g., the token-by-token generation process.
    The format should clearly separate placeholder values, avoid conflicts with surrounding code, and remain interpretable by LLMs.
    
    \item \textbf{Which code patterns should be sugarized?}  
    Not all code patterns are equally valuable for sugarization.
    Manually selecting patterns based on experience is neither scalable nor guaranteed to yield optimal token savings. Instead, we require a systematic, data-driven approach to identify frequently used, token-heavy patterns that offer significant compression potential.

    \item \textbf{How to teach LLMs to utilize token sugars effectively?}  
    The shorthand in token sugars introduces new concepts into the language space of the LLMs.
    To ensure fluent usage, LLMs require specialized training on sugarized examples
    It is still unclear how to perform such training and what training strategies are needed.

\end{itemize}

In the following sections, we will demonstrate how these challenges can be addressed through a proof-of-concept design and implementation.

\begin{figure*}[!t]
    \centering
    \includegraphics[width=\textwidth]{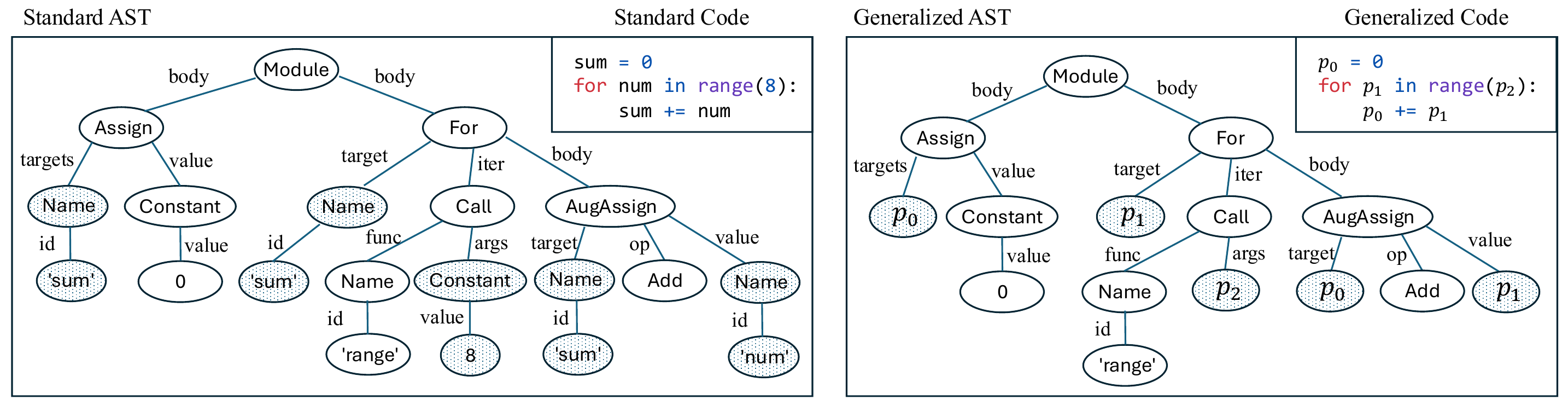}
    \caption{An example of how a standard AST is generalized. Shadowed nodes are replaced during generalization.}
    \label{fig:ast}
\end{figure*}

\section{Methodology}
This section explains how to apply token sugar to facilitate the LLM-based code generation process.
We focus on Python as a proof of concept, considering its wide support in existing LLMs and its reputation for conciseness~\cite{bergmans2021measuring}.
If Token Sugar achieves significant token savings in Python, it is likely to generalize well to more verbose languages.

Specifically, our methodology consists of three components, each corresponding to a core design challenge:
(1) a shorthand that optimizes token efficiency while ensuring reversibility,
(2) a data-driven mining pipeline for discovering high-value code patterns for sugarization, and
(3) a training strategy integrating token sugars into LLMs via sugarizing the training corpus. 
In the subsequent sections, we introduce each component in detail.

\subsection{Shorthand Design}~\label{sec:shorthand-format}
The design space for a shorthand system is vast.
As the first attempt, we propose a heuristic design that aims to satisfy two core requirements: efficient token usage and deterministic reversibility.
While heuristic in nature, this design serves as a viable proof of concept—if effective, it paves the way for more advanced shorthand schemes in the future.
Since a code pattern typically appears within a larger code context, its shorthand replacement thus coexists with the surrounding code.
To allow the converter to correctly recognize the shorthand in any code context, its design needs to eliminate the ambiguity with the standard Python grammar.
Specifically, we first introduce special tokens that do not exist in Python grammar into the shorthand representation and then use additional or existing marks to clearly indicate the scope of the shorthand.

Each special token is associated with a unique code pattern ID.
We represent such a token as $\langle ID \rangle$ in this paper.
These special tokens will be introduced into the LLM tokenizer's vocabulary and thus treated as a single atomic token (i.e., it will not be further tokenized into multiple sub-tokens).
When the converter recognizes such a token in its given sugarized code, it will use the ID to retrieve the corresponding transformation rule for further desugarization.
To indicate the scope of the shorthand, we differentiate the shorthand format based on the type of code pattern: \textit{statements} versus \textit{expressions}.

\subsubsection{Shorthand for Statement-Level Patterns}

In Python grammar~\cite{Python_Full_Grammar_Specification}, a \textit{statement} represents 
a complete instruction that can be executed, such as assignments 
(e.g., `x = 5'), control flow constructs (e.g., `for' loops, `if' statements), 
or function/class definitions.
Python uses newline characters after statements, which naturally separates statement-level patterns from neighboring code.
As such, we can directly use newlines at the start and end of the shorthand to indicate the scope.
We design an inline-delimited format that separates two groups of placeholders in the code patterns.
The left-hand side (LHS) group includes placeholders corresponding to variables whose values are written (i.e., variables that are newly created and assigned) in the pattern.
For example, the target variable of an assignment statement (e.g., $p_0$ in Fig. 2) and the loop variable in a for loop (e.g., $p_1$ in Fig. 2), 
All the remaining placeholders in the code pattern are included in the right-hand side (RHS) group.
Placeholders in both groups are ordered based on their appearance in the pattern.

Formally, let $p_0, p_1, \dots, p_x$ denote LHS placeholders, and $p_{x+1}, \dots, p_n$ denote RHS placeholders.
The shorthand is constructed as follows, separating two groups using a special token $\langle ID \rangle$, and a delimiter character ``$;$'' within each group:
\[
\underbrace{p_0 \, ; \, p_1 \, ; \, \cdots \, ; \, p_x}_{\text{LHS placeholders}}
\underbrace{\langle ID \rangle}_{\text{Token}}
\underbrace{p_{x+1} \, ; \, \cdots \, ; \, p_n}_{\text{RHS placeholders}}
\]
For instance, the common Python pattern ``for $p_0$ in range($p_1$): $p_2$.append($p_3$)" can be represented as "$p_0$$\langle1001\rangle$$p_1;p_2;p_3$''.
The placeholder $p_0$ is on the LHS because it is the loop variable defined in this `for' statement; $p_1$, $p_2$, and $p_3$ are defined outside the statement.
By using the special token as a delimiter between the two groups, we can often avoid the need for one additional separator token (i.e., the semicolon we use within each group), thus saving one more token in the final shorthand representation.
The LHS/RHS groups are also defined to logically align with standard assignment operations, where new variables (LHS) are defined/updated based on existing ones (RHS).

\subsubsection{Shorthand for Expression-Level Patterns}
In Python grammar, \textit{expression-level} patterns can appear within other code structures, arithmetic operations (e.g., `x + 1'), 
function calls (e.g., `len(s)'), list comprehensions (e.g., `[x**2 for x in range(10)]'), 
and conditional expressions (e.g., `x if x $>$ 0 else -x'), making their boundaries less obvious than statement-level patterns.
To ensure deterministic reversibility, we adopt a wrapped format that explicitly marks the start and end of the shorthand using special tokens:
\[
\underbrace{\langle ID\rangle}_{\text{Token}} 
\underbrace{p_0 \, ; \, p_1 \, ; \, \cdots \, ; \, p_n}_{\text{Placeholders}}
\underbrace{\langle END\rangle}_{\text{Token}}
\]
Here, we do not need to separate placeholders into LHS and RHS groups, as Python does not use expressions to define variables.
Compared to the statement-level shorthand, this structure introduces an end token $\langle END\rangle$ to explicitly delimit the shorthand within arbitrary contexts.
For example, the arithmetic pattern ``$p_0$ * ($p_0$ + 1)" becomes ``$\langle 1002\rangle p_0\langle END\rangle$''.
A code context embedded with this shorthand, such as ``x =  $\langle 1002\rangle p_0\langle END\rangle$ + 1'', can be reversed to the original Python code without ambiguity.

\begin{figure*}[!t]
    \centering
    \includegraphics[width=0.85\textwidth]{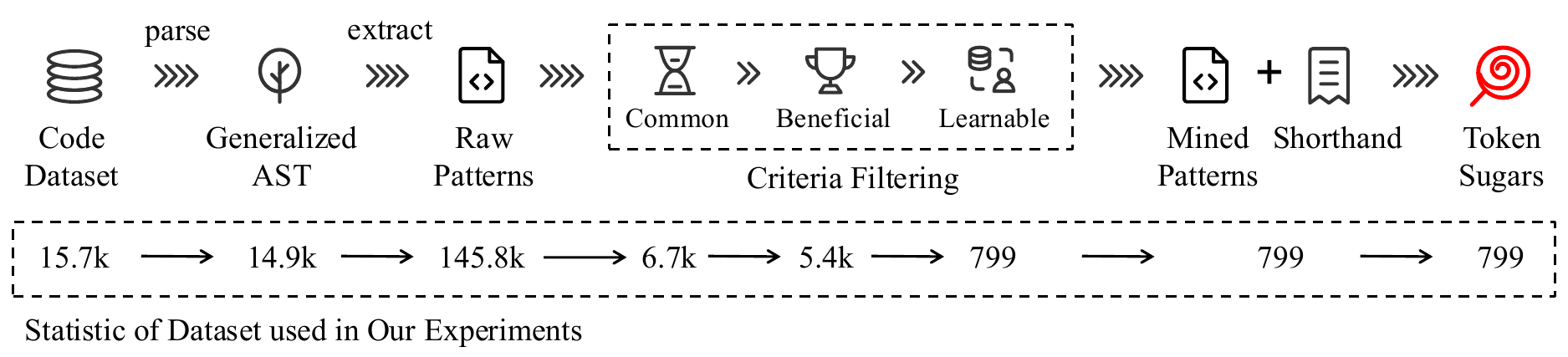}
    \caption{An overview of the mining pipeline.}
    \label{fig:mining-pipeline}
\end{figure*}

\subsection{Sugarizable Code Pattern Mining}~\label{sec:pattern-mining}
While individual token sugars can optimize specific code patterns, their isolated impact on overall token reduction is limited due to the diversity and complexity of real-world coding tasks.
To meaningfully reduce token usage at scale, we must identify a large and diverse set of code patterns, from which token sugers are derived.
Given the large decision space, manual selection is impractical, motivating the need for an automated, data-driven approach to discover high-value sugarization candidates across diverse coding contexts.
Therefore, we propose a data-driven mining pipeline that systematically identifies recurrent, token-heavy code patterns suitable for sugarization.
As illustrated in~\Cref{fig:mining-pipeline}, this process transforms raw source code into generalized abstract syntax trees (ASTs), applies frequent subgraph mining, and finally filters candidates based on their token-saving potential.

\subsubsection{Generalized Abstract Syntax Tree}\label{sec:ast}
Our mining task shares some similarity with prior work on mining code idioms~\cite{allamanis2014mining,obrien2024data,sivaraman2022mining,yang2024streamlining}.
While prior studies focus on idioms suitable for human-oriented syntactic sugar or code reuse, our goal is to identify patterns that maximize token savings for LLMs.
Such distinctions necessitate a tailored mining pipeline.
For example, prior work on mining code idioms~\cite{obrien2024data, yang2024streamlining} mines from control flow graphs (CFGs).
While CFG-based mining method is effective for code reuse, it abstracts away syntactic details by design.
For example, in CFG, loops are often normalized into equivalent branch-and-backedge structures, where either a for-loop or a while-loop is the same.
For LLM token savings, these distinctions are essential: a for loop may consume fewer tokens than a while loop for the same logic.
To preserve these token-sensitive details, we mine from ASTs instead of CFGs.

As found in~\cite{allamanis2014mining,obrien2024data}, mining frequent subgraphs directly on raw graphs parsed from code corpus results in very small and meaningless idioms.
Restriction to the mining scope is thus necessary to filter out information that is not interesting and also reduce the search space.
For example, to guide the design of syntactic sugar, OBrien et al.~\cite{obrien2024data} adopt a generalized context-free grammar (CFG) that abstracts away noise, such as project-specific information.
Similarly, we generalize the abstract syntax tree (AST) by abstracting away some developer-defined elements that are unlikely to be valuable for sugarization.
Specifically, we choose to normalize the nodes for variable names and uncommon constants and replace them with generic placeholder nodes, ensuring that each unique variable or constant retains a consistent placeholder throughout the AST.
An shown in~\Cref{fig:ast}, identifiers like ``sum'' and ``num'' and the constant ``8'' are generalized to $p_0$, $p_1$, and $p_2$, respectively. However, we preserve a curated set of high-frequency constants including numeric literals (0, 1, -1, 2, 3, 10), boolean values (True, False), None, and empty strings, due to their prevalence as semantic markers in Python source code.
Based on our analysis through a large-scale Github code corpus, starcoderdata~\cite{li2023starcoder}, these constants are the top 10 most frequent ones in Python.
From such a generalized AST, we mine the code patterns.

\subsubsection{Mining Pipeline}
The goal of our mining is to find the valuable code patterns that can be sugarized to reduce the token usage of source code.
This can be translated to three requirements:
\begin{itemize}[leftmargin=*]
    \item \textit{Common Usage}: the code pattern should be commonly seen or used by LLMs.
    \item \textit{Saving Potential}: the code pattern should reduce token savings if sugarized.
    \item \textit{Training Data Availability}: the code pattern should have sufficient training data for LLMs to learn.
\end{itemize}
Guided by these requirements, we propose the mining pipeline, where an overview is shown in~\Cref{fig:mining-pipeline}.
Specifically, given a code corpus, we parse the code samples into their generalized ASTs, from which we extract all possible code patterns and filter them with a series of rules.
In the following, we detail the pipeline of mining code patterns that satisfy the above criteria.

\smallskip\noindent \textbf{Dataset Selection}:
Straightforwardly, the code patterns should be mined from the training data of LLMs.
However, the training dataset of LLMs, such as source code in Github repositories, is quite different from the tasks given by the users of an LLM.
For example, boilerplate code like ``if \_\_name\_\_ == '\_\_main\_\_':'' is common in training data but irrelevant for assistant-style LLM use cases.
These irrelevant patterns can obscure the code structures that truly matter for practical use cases.
To ensure relevance, we prioritize in-distribution datasets that reflect actual user scenarios.
The gold standard would be production data from deployed LLM applications, as it directly captures real-world coding patterns.
When such data is unavailable (e.g., for LLM developers without access to deployment logs), we can use proxy datasets that share similarity to LLM's practical use cases.
For example, in our experiment, we use the code dataset collected from Leetcode to serve as a proxy in-distribution dataset for the function-level code generation tasks.

\smallskip\noindent \textbf{Raw Patterns Extraction}
Given the dataset to be mined, we first parse its code samples into generalized ASTs, as described in~\Cref{sec:ast}.
Each generalized AST is then decomposed into connected subtrees, with three structural constraints to reduce noisy patterns: 1) single node: complete statement or expression nodes in Python; 2) adjacent nodes: contiguous node sequences in source code; 3) node heads: compound statement nodes (e.g., \texttt{for}, \texttt{while}, \texttt{if}) without their body sub-node.
For candidate patterns yielded from this decomposition, we count their occurrences in the dataset and only keep the ones that appear in more than $k$ data samples.
This frequency threshold filters out project-specific or rare patterns, focusing on widely-used ones.

\smallskip\noindent \textbf{Estimating Saving Potential}:
Not all code patterns benefit from sugarization since some are already concise.
To prioritize high-impact candidates, we estimate each pattern's token-saving potential by comparing token counts between original patterns and their shorthand equivalents.
Specifically, we instantiate both forms using the same variable names and measure their token lengths when processed by the target LLM's tokenizer. 
Patterns demonstrating savings below a threshold of $m$ tokens are discarded.

\smallskip\noindent \textbf{Checking Training Data Availability}:
For LLMs to effectively learn the shorthand representations, the mined code patterns should be supported by sufficient training data, i.e., enough training samples can be matched by the code pattern.
We address this by filtering candidates based on their frequency in the target LLM's training corpus. 
Specifically, we retain only those patterns that appear at least $n$ times.

The remaining code patterns are converted into token sugars through a systematic process.
Each pattern is assigned a unique identifier (ID) and coupled with its corresponding shorthand following the format defined in~\Cref{sec:shorthand-format}.
These pattern-shorthand mappings are stored in a structured lookup table, enabling efficient retrieval during code transformation operations using the pattern ID as a key. 
This bidirectional mapping system facilitates seamless conversion between standard and shorthand code representations, serving the sugarization of training data for model learning and the transformation as either a pre-processing or post-processing step in LLM inference pipelines.

\subsection{Training Dataset Construction}~\label{sec:training-strategy}\label{sec:dataset-construction}
With a curated set of token sugars derived from our mining pipeline, the next step is to enable LLMs to utilize them during inference.
Rather than altering the model architecture or training algorithm, our approach integrates token sugars directly into the training data.
This strategy ensures compatibility with standard training workflows.

Specifically, given a training dataset, we derive a sugarized dataset by transforming code samples captured by the code patterns to the corresponding shorthand.
During this process, conflicts arise when multiple token sugars overlap within a single code region.
For example, one pattern may span lines 1–2 while another matches lines 2–3, creating a conflict regarding which transformation to apply.
To resolve such conflicts systematically, we model the problem as a weighted interval scheduling task~\cite{kolen2007interval}, where each sugarizable pattern match is treated as an interval. 
The interval's range corresponds to the code lines it spans, and its weight represents the estimated token savings derived during the Estimating Saving Potential step of our mining pipeline.

To solve this task, we adopt a greedy optimization strategy that seeks to select a conflict-free subset of transformations to maximize total token savings.
Specifically, we first sort all token sugar applications by their end line number, ensuring that we can apply dynamic programming over increasing end positions.
For each application $i$, we determine $p(i)$, the latest non-overlapping application that ends before $i$ starts.
This can be computed efficiently using binary search over the sorted list.

We then define a dynamic programming recurrence:
\[
\text{dp}[i] = \max(\text{dp}[i-1],\ \text{dp}[p(i)] + w_i)
\]
where $\text{dp}[i]$ denotes the maximum token savings achievable using the first $i$ token sugar matches, and $w_i$ is the estimated token saving of the $i$-th match.
The term $\text{dp}[i-1]$ corresponds to skipping the current match, while $\text{dp}[p(i)] + w_i$ corresponds to selecting it along with the optimal set of non-overlapping matches before it.
We populate the $\text{dp}$ table from $1$ to $n$, where $n$ is the total number of candidate matches.
The final value $\text{dp}[n]$ gives the maximum total token savings.
Once the table is computed, we perform a standard traceback to retrieve the final set of selected, non-overlapping matches.

Using the selected non-overlapping matches, we rewrite the code samples by replacing each matched pattern with its corresponding shorthand.
This results in a sugarized training dataset that is both conflict-free and optimized for token efficiency.
To preserve the model’s ability to understand natural, unsugarized code, we adopt a strategy inspired by methods for mitigating catastrophic forgetting~\cite{zheng2024breaking}.
Specifically, we retain 25\% of the original training samples in their unmodified form.
This mix ensures the model remains fluent in conventional code while progressively learning to interpret and generate the sugarized form.
Finally, to ensure correct processing of the shorthand during training, all special tokens of the mined token sugars are explicitly added to the tokenizer’s vocabulary. 
With these preparations, the resulting dataset can be used to train the LLM.
The model thus acquires the ability to seamlessly understand and produce token sugar representations, effectively bridging human-readable code and token-efficient abstractions.
\section{Experiment Setup}
Using our proposed method as a demonstration, we experimentally evaluate the feasibility of Token Sugar.
In this section, we will introduce the settings of the experiments, including the research questions, datasets and corresponding mined token sugars, evaluation metrics, and the models used in the experiments.
The rationale behind our setup is driven by answering the following two research questions:
\begin{itemize}[leftmargin=*]
    \item \textbf{RQ1}: What is the token reduction capability of mined token sugar in representing source code?
    \item \textbf{RQ2}: Can adapted LLMs effectively utilize mined token sugars, while maintaining their coding capability?
\end{itemize}

\subsection{Datasets}
Due to the limited availability of the production data, we have to adopt a simulated scenario so that we can get proper in-distribution code samples as the proxy of the production data.
To be specific, we assume that in the experiments, we are developing LLMs as coding assistants focusing on the function-level code generation tasks.
Therefore, two kinds of datasets are involved: the training dataset for LLMs and the in-distribution dataset for mining token sugars. 

For the training data of LLMs, we utilize the Python subset of starcoderdata~\cite{li2023starcoder}, a filtered variant of The Stack dataset~\cite{kocetkov2022stack}.
It is a large-scale code corpus for LLM pre-training, containing over 20 million code files sourced from open-source GitHub repositories.
We keep the code files from the repositories with over 100 stars, resulting in 623,887 code files.
Notably, this dataset has been filtered to the data samples in the evaluation benchmark in our experiments.

For the function-level code generation task in our experiments, we use Leetcode solutions~\cite{LimYeri_LeetCode_Python_Solutions_v2} to serve as the proxy for the code that LLMs are required to generate in practice.
This choice is motivated by the fact that Leetcode solutions primarily consist of function-level implementations, mirroring the coding patterns commonly encountered in function implementation, such as parameter validation and return value handling.
This dataset, containing 15,734 code samples, is collected from the Python solutions to coding problems that have received at least 10 votes on the Leetcode platform.
We also performed string matching against our evaluation benchmark to prevent potential data leakage.

\subsection{Mined Token Sugars}
Following our proposed method, we mine token sugars from the LeetCode solutions.
During the mining process, we heuristically set the threshold $k$ for common code patterns to 5, i.e., each code pattern must be used by at least 5 different solutions.
We also set the minimum token saving of a code pattern $m$ to 1, ensuring each token sugar can achieve positive token gain.
For the training data availability, we refer to experiences in studies on data poisoning, where manipulating 0.1\% of the training data is sufficient to change the model's behavior~\cite{sun2024fdi}, and set the minimum appearance of each code pattern in the training data to 0.1\% of the total number of code files, i.e., 623.
This set of parameters finally results in 799 available token sugars.
With these token sugars, we implemented a prototype converter that enables the sugarization and desugarization process, which will be used for our experiments, i.e., sugarizing the training dataset and desugarizing the LLM-generated code.
The statistics during the mining process are reported at the bottom of~\Cref{fig:mining-pipeline}.

\subsection{Evaluation Metrics}
We evaluate the model's performance on the function-level code generation task with the Pass@1 metric on HumanEval~\cite{chen2021evaluating}.
To compute Pass@1, one code samples are generated for each problem, and a problem is considered solved if the generated code passes the unit tests.
We report the fraction of problems being successfully solved.
The HumanEval dataset comprises 164 programming problems, each with a function signature, a docstring, and multiple test cases.
Given the function signature and docstring, the model is required to generate the code, which is then tested by executing the test cases.
Notably, for the model adapted to token sugars, we desugarize the generated code to run test cases.

To analyze the token sugar utilization during generation, we measure three additional metrics: \%Saved Tokens, \#Sugars, and \#Failed. 
\%Saved Tokens denotes the percentage of tokens saved by employing LLMs with token sugars.
Direct token count comparisons between two models' generated code for the same task can be misleading due to the different implementations they may choose.
Thus, we estimate the \%Saved Tokens by desugaring the generated code from the model and comparing the token counts before and after this process.
\#Sugars denotes the average number of sugars used per generated sample.
\#Failed captures the number of cases where the generated token sugar cannot be successfully desugared.
This metric helps assess whether an LLM has correctly learned how to use token sugars.
For instance, if a shorthand defined with a single LHS placeholder is incorrectly used with two LHS variables by LLMs, the desugaring process will fail.

\subsection{Model}
In our method, the mined token sugars are learned through continual pre-training.
We adopt three popular LLMs to serve as the initial model for our experiments, including Pythia, Llama-3.2, and Qwen-2.5.

\begin{itemize}[leftmargin=*]
    \item \textbf{Pythia}: Pythia~\cite{biderman2023pythia} is a suite of LLMs designed close to currently accepted common practices. We use its 1.4B version, the largest one in the suite that satisfies our resource constraint.
    \item \textbf{Llama-3.2}: Llama-3.2~\cite{MetaAI_Llama3_Connect_2024} is a newly released LLM by Meta, which outperforms many of the available open source and closed LLMs on common industry benchmarks. Considering our resource constraint, we use its 1B version.
    \item \textbf{Qwen-2.5}: Qwen-2.5~\cite{qwen2024qwen205} is an enhanced version of Alibaba's Qwen large language models. We use its 1.5B version for our experiments.
\end{itemize}
Though their size is relatively small, they suffice to validate the feasibility of learning token sugars.
We will further discuss the impact of this decision in the discussion section.

\subsection{Implementation Details}
In our experiments, we use the Huggingface Transformers~\cite{wolf2020transformers} library with PyTorch to implement the models.
The models are trained on a machine with 128 vCPUs, 200GB RAM, and four RTX A6000 GPUs (48GB RAM).
For all the three models, we adopt a batch size 12 with 64 accumulation steps and 512 context length. 
The learning rate is set to 1.8e-4 with a cosine decay schedule.
During inference, we use the greedy decoding strategy and a maximum of 512 tokens are generated.

\section{Results}
This section reports the experimental results and answers the research questions.

\subsection{RQ1: Token Reduction}

\begin{table}[!t]
    \centering
    \caption{Token saving results.}
    \begin{tblr}{
    cells = {c},
    colsep = 3.5pt,
    cell{1}{1} = {r=2}{},
    cell{1}{2} = {c=7}{},
    cell{2}{3} = {c=2}{},
    cell{2}{5} = {c=2}{},
    cell{2}{7} = {c=2}{},
    vline{1-2,9} = {1-2}{},
    vline{3,5,7} = {2}{},
    vline{1-3,5,7,9} = {3-4}{},
    hline{1,3-5} = {-}{},
    hline{2} = {2-8}{},
  }
  \textbf{Dataset} & \textbf{\#Tokens} &  &  &  &  &  & \\
   & \textbf{Original} & \textbf{SimPy} &  & \textbf{Token Sugar~} &  & \textbf{Combined} & \\
  LeetCode & 2.0m & 1.7m & 15.3\%$\downarrow $ & 1.7m & 15.1\%$\downarrow $ & 1.5m & 22.4\%$\downarrow $\\
  Humaneval & 12.8k & 11.1k & 13.3\%$\downarrow $ & 11.3k & 12.9\%$\downarrow $ & 10.2k & 20.0\%$\downarrow $
  \end{tblr}
    \label{tab:token_saving}
\end{table}

For RQ1, we quantified the token reduction achieved by Token Sugar across two distinct code sets: our in-distribution dataset (Leetcode solutions) and the HumanEval ground-truth code, which represents the code that LLMs are expected to generate for function-level code generation. 
It allows us to assess the token reduction capability of Token Sugar in relevant scenarios.
Since Token Sugar is proposed to capture the unrealized token reduction potential beyond what is achievable by existing syntax-based methods, we further examined its complementarity with SimPy~\cite{sun2024ai}, a state-of-the-art syntax-level code simplification method.
To enable a comprehensive analysis, we defined four experimental groups:: \textbf{Original} (our baseline, code without any processing); \textbf{Token Sugar} (code processed solely by token sugars); \textbf{SimPy} (code processed solely by SimPy); and finally, \textbf{Combined} (code processed by both Token Sugar and SimPy).
For the \textbf{Combined} group, we applied Token Sugar and SimPy together to the code, where SimPy was modified to be compatible with Token Sugar.

Our experimental results demonstrate that Token Sugar can achieve significant token savings.
As shown in~\Cref{tab:token_saving}, Token Sugar alone achieves a token reduction of 15.1\% on LeetCode and 12.9\% on HumanEval.
This demonstrates that Token Sugar, as a lossless method, effectively reduces the number of tokens required to represent code without altering its semantics.
More importantly, when used in conjunction with SimPy (which individually achieves 15.3\% and 13.3\% reductions), Token Sugar brings additional savings, pushing the total reduction to 22.4\% on LeetCode and 20.0\% on HumanEval.
These results validate Token Sugar’s core premise: addressing verbosity that remains untouched by purely syntactic simplifications.
While SimPy reduces tokens by optimizing syntax rules (e.g., indentation, delimiters), Token Sugar targets semantically repetitive patterns (e.g., API calls, loop structures, numeric literals).
This reveals the unique role of Token Sugar in complementing existing techniques to address the token waste.

\begin{tcolorbox}[size=title]
    {\textbf{Answer to RQ1:}} Token Sugar achieves significant token reduction in Python code (12.9–15.1\%) while targeting different verbosity sources, and the combined use with SimPy unlocks synergistic reductions (20.0–22.4\%).
\end{tcolorbox}

\subsection{RQ2: Model Performance}
To answer RQ2, we evaluate how token sugars are utilized when the trained LLMs generate code and whether training with a sugarized dataset affects the model's code generation capability.
Specifically, we perform further pre-training on three base models (Pythia, Llama-3.2, and Qwen-2.5) using both sugarized and original training data, respectively.
We then evaluate the performance of the trained models on the HumanEval dataset using the Pass@1 metric.
The outputs generated in this process are further analyzed to compute the three metrics previously introduced: \%Saved Tokens, \#Sugars, and \#Failed.

\begin{table}[!t]
    \centering
    \caption{Model performance results.}
    \begin{tblr}{
    colsep = 3.5pt,
    cells = {c},
    cell{1}{1} = {r=2}{},
    cell{1}{3} = {c=4}{},
    vline{1-3,7} = {1-5}{},
    vline{4,5,6} = {2-5}{},
    hline{1,3-6} = {-}{},
    hline{2} = {2-6}{},
  }
  \textbf{Model} & \textbf{Baseline} & \textbf{Token Sugar} &  &  & \\
   & \textbf{Pass@1} & \textbf{Pass@1} & \textbf{\%Saved Tokens} & \textbf{\#Sugars} & \textbf{\#Failed}\\
  Pythia & 6.7\% & 6.7\% & 7.7\% & 2.3 & 0\\
  Llama-3.2 & 12.8\% & 12.2\% & 8.3\% & 1.2 & 0 \\
  Qwen-2.5 & 26.2\% & 25.6\% & 11.2\% & 2.5 & 0
\end{tblr}
    \label{tab:model_performance}
\end{table}

As shown in~\Cref{tab:model_performance}, the Pass@1 scores remain remarkably stable when comparing models trained on the dataset before and after sugariation.
Specifically, Pythia maintains identical performance (6.7\%), while Llama-3.2 and Qwen-2.5 show only marginal differences of -0.6\% and -0.6\% respectively.
This minimal performance variance demonstrates that Token Sugar preserves the essential semantic information needed for code generation tasks despite its token-saving benefits.
Diving into the usage of token sugars, we observe a consistent level of token savings across models, with \textbf{Qwen-2.5} achieving the highest savings at 11.2\%, followed by \textbf{Llama-3.2} (8.3\%) and \textbf{Pythia} (7.7\%).
These results indicate that stronger models, i.e., models with higher Pass@1, tend to leverage token sugars more effectively, likely due to their enhanced capacity to learn and apply abstracted patterns during generation.
Also, these values are slightly lower than the token reduction measured in RQ1, 15.1\% on Leetcode and 12.9\% on HumanEval, but remain substantial.
This discrepancy is expected, as the model is not forced to use token sugars and must autonomously learn when and how to apply them during generation.
The observed savings confirm that the token sugars mined can be practically adopted by LLMs in code generation tasks.

Moreover, the average number of sugars used per generated code sample (\#Sugars) does not show a perfect correlation with their accuracy reflected through Pass@1.
Qwen-2.5, the model with highest Pass@1, uses the most sugars (an average of 2.5 token sugars per generated code), and achieves the best token saving (11.2\%), suggesting effective and frequent use of token sugars.
Interestingly, Llama-3.2 uses the fewest sugars on average (1.2 token sugars per generated code), yet achieves a greater token reduction (8.3\%) than Pythia, which uses more sugars (2.3 token sugars per generated code) but saves slightly less (7.7\%).
This suggests that Pythia may overuse sugars without achieving proportionate savings, possibly due to its less precise understanding of token sugars.
Finally, across all models, we observe a failure count of zero, indicating that LLMs understand token sugars well and can generate them in a valid format, i.e., all the required formats, including the number and position of placeholders and delimiters.
This is critical for the usability of token sugars in real applications.

\begin{tcolorbox}[size=title]
    {\textbf{Answer to RQ2:}} Token sugars are actively and reliably used by LLMs during code generation, causing a negligible difference in Pass@1 performance while yielding up to 11.2\% token savings without any desugarization failures. Stronger models show more strategic and efficient use of token sugars.
\end{tcolorbox}

\section{Related Work}

\subsection{Program Simplification}
Program simplification has been widely explored as a means to improve the efficiency and interpretability of neural code models~\cite{Huang2023ProgramTV,Rabin2022SyntaxguidedPR,Rabin2021UnderstandingNC,Shi2023StructuralsemanticsGP,Bui2019AutoFocusIA,Yang2024RobustnessSP,Zheng2020ProbingMS,Zhang2022DietCI,shenempirical,wang2025leancode,pan2025hidden}.
These methods typically reduce input length by removing or rewriting less important code tokens, aiming to reduce the cost of inference or training.
For example, DietCode\cite{Zhang2022DietCI} simplifies code by removing code tokens that receive the lowest attention scores from CodeBERT.
While effective in improving efficiency or interpretability, these approaches are inherently lossy, removing or altering information that may be necessary for recovering the original code.
Therefore, such methods are limited to code understanding tasks like code summarization and retrieval.
To address the limitations of lossy simplification, recent work has explored reversible simplification through grammar-level transformations.
SimPy~\cite{sun2024ai}, for instance, proposes using an AI-oriented syntax to reformat code in a way that preserves semantics while improving model efficiency.
However, this approach is confined to syntactic transformations and does not reduce deeper semantic redundancies that arise from repetitive logic or idiomatic patterns.
In this paper, we fill this gap by proposing Token Sugar, a reversible program simplification method at the semantic level.

\subsection{Code Idiom Mining}
Code idiom mining~\cite{allamanis2018mining} focuses on identifying recurring code patterns that encapsulate specific semantic roles.
Usually, such approaches focus on mining code idioms to assist human developers by promoting code reuse, improving readability, and informing language design~\cite{iyer2019learning, allamanis2018mining, shetty2023codescholar,yang2024streamlining, obrien2024data,sivaraman2022mining, grand2023learning, cao2023babble}.
For example, IdioMine~\cite{yang2024streamlining} automatically extracts well-formed idioms from Java projects to streamline programming practices.
O'Brien et al.~\cite{obrien2024data} propose a data-driven approach to syntactic sugar design by mining frequent subgraphs in Java methods, aiming to simplify common code idioms.
Since AI has become prevalent in code-related tasks, some studies~\cite{shin2019program, iyer2019learning, stengel2024regal} also explore code idioms from a model-centric perspective.
Among these studies, the works by Iyer et al.\cite{iyer2019learning} and Shin et al.\cite{shin2019program} demonstrate that code idioms can improve the efficiency of AI models in the code generation task.
However, their methods are primarily designed for traditional Seq2Seq models with grammar-rule-based decoding, where code generation is framed as a step-by-step construction of ASTs, typically by adding one sub-node or edge at a time.
In contrast, modern LLMs, such as Deepseek~\cite{liu2024deepseek}, extend a linear sequence of tokens one by one, without explicit structural constraints.
This paradigm shift raises new challenges and opportunities for leveraging code idioms in LLMs, particularly in terms of how idioms can be learned, represented, and utilized effectively.
\section{Discussion}\label{sec:discussion}

\subsection{Limitations}
\noindent \textbf{Constrained Model Size}
The model selection in our experiments is restricted by our computational resources, limiting our evaluation to models with around 1B parameters.
These models are relatively modest in scale.
However, while the model size is expanding, the fundamental issue of computation waste caused by redundant source code elements remains unaddressed.
Therefore, the insights derived from our experiments with smaller models are still highly relevant for understanding inefficiency issues in larger models.

\noindent \textbf{Standalone Token Sugar Mining}
In RQ1, we reveal that Token Sugar achieves significant token savings, both as a standalone method and when combined with existing techniques like SimPy.
However, the token sugars used in our experiments were mined from the original version of the code, not from the SimPy-processed variant.
This may lead to overlapping or redundant effects when combined with SimPy, thereby resulting in an underestimation of Token Sugar's contribution.

\noindent \textbf{Python Focus}
Similar to SimPy, our research primarily realizes Token Sugar in Python, a popular and concise programming language.
It has successfully revealed the potential and feasibility of this general concept.
Since programming languages differ in syntax, verbosity, and structural conventions, our realization may not directly transfer to other languages.
Further work to implement Token Sugar in multiple languages is feasible and may offer even greater opportunities for token reduction.

\subsection{Overhead for Desugaring Token Sugars}
An essential part of our approach is the desugarization process, where token sugars in the generated code are expanded back to their original form for execution and human readability.
While introducing a transformation step may raise concerns about additional computational overhead, we find that in practice, this cost is negligible.
Our experimental implementation uses an unoptimized Python script to perform desugarization.
Even under this baseline setup, the average time to process a single token sugar is 1.3 milliseconds, a comparable speed to SimPy's code converter.
In production environments, this step can be further optimized through more efficient implementations or integration into the model-serving pipeline.
Therefore, the cost of desugarization is not a practical barrier to deployment.

\subsection{Access to Production Data}
Our method requires access to the code usage data of LLMs, or data that closely approximates it, to effectively mine token sugars.
When such data is unavailable due to privacy, legal, or infrastructural constraints, the method may not be able to achieve optimal token savings.
However, it is important to emphasize that the primary beneficiaries and intended users of Token Sugar are platforms that provide LLM-based code generation services.
These platforms are typically in a unique position: they possess both the infrastructure to train large models and continuous access to vast amounts of source code generated or seen by LLMs, collected through interactions with their deployed coding assistants~\cite{sun2024fdi}. 
For example, according to their official documents~\cite{GitHub_Copilot_Features,Cursor_Privacy_Policy,AWS_CodeWhisperer_Data_Sharing}, Github Copilot, Cursor, and Amazon CodeWhisperer are collecting data from permissive users.
More importantly, they face substantial computational costs at inference time and have strong motivations to reduce them.
Therefore, while our method assumes access to production-like data, this is not a practical limitation for its core audience.

\subsection{On the ``Sweet Spot" of Token Sugar}
A significant open question arising from our work is determining the optimal balance, or ``sweet spot", for the hyper-parameters of the Token Sugar approach, such as the filtering thresholds.
The research presented in this paper serves as a foundational proof-of-concept, which seeks to first establish the feasibility of introducing token-efficient constructs and, more importantly, to demonstrate that LLMs can learn them without a significant degradation in performance.
To this end, we employed several heuristic-based thresholds as a practical starting point for our initial investigation.
However, we acknowledge that a trade-off exists.
Simply increasing the number of Token Sugars will not linearly improve performance.
Eventually, this will lead to diminishing returns as the model's capacity to learn these new patterns becomes saturated.
A systematic exploration of this solution space to identify the optimal application strategy is an exciting and essential direction for future work.

\subsection{Generalization to vanilla LLMs}
By design, our constructs are out-of-distribution for LLMs that are not explicitly trained on token sugars.
We conducted two experiments using GPT-4.1 on a code completion task to investigate this scenario.
Specifically, we split the standard solution of Humaneval into two halves (prefix and suffix) and instruct GPT-4.1 to complete the prefix.
When provided with a standard Python prefix, GPT-4.1 achieved a baseline Pass@1 of 94.5\%.
However, when we sugarized the prefix and presented it to the same model, the Pass@1 score dropped dramatically to 51.2\%.
Next, we ran the same experiment but included examples of Token Sugars and their expanded forms within the prompt, resulting in a Pass@1 score of 54.9\%.
These findings underscore a key aspect of our approach: unlocking the full potential of Token Sugar requires dedicated model training.

\section{Conclusion and Future Work}
In this paper, we introduce the concept of \textit{Token Sugar}, an abstraction mechanism aimed at reducing the verbosity of programming languages for LLMs.
We further realized this idea with a systematic attempt, which successfully extracted 799 token sugars and integrated them into three mainstream LLMs via a sugarized training dataset.
Through a series of empirical studies guided by practical implementation and evaluation, we demonstrate the feasibility, scalability, and effectiveness of Token Sugar.

As a newly proposed direction, Token Sugar opens up many unanswered questions that extend beyond the scope of our current proof-of-concept implementation.
For instance, training LLMs to understand and generate token sugars through continual pretraining can be resource-intensive.
Future work could explore more efficient approaches, such as parameter-efficient fine-tuning methods, to reduce the computational cost.
We encourage the research community to further investigate this promising area.

\section*{acknowledgments}
This research / project is supported by Xiaoning Du’s Google Research Scholar Program Award and the National Research Foundation, under its Investigatorship Grant (NRF-NRFI08-2022-0002).
Any opinions, findings and conclusions or recommendations expressed in this material are those of the author(s) and do not reflect the views of National Research Foundation, Singapore.

\balance
\bibliographystyle{IEEEtran}
\bibliography{IEEEexample}

\end{document}